\documentclass[%
reprint,
superscriptaddress,
amsmath,amssymb,
prb
]{revtex4-2}

\usepackage{graphicx}
\usepackage{dcolumn}
\usepackage{float}
\usepackage{bm}% bold math

\begin{document}
\title{$^{13}$C NMR observation of a nonmagnetic charge-ordered state in the organic conductor $\kappa$-(ET)$_{2}$Hg(SCN)$_{2}$Cl}
\author{Mizuki Urai}
\email{urai@issp.u-tokyo.ac.jp}
\affiliation{Department of Applied Physics, University of Tokyo, Tokyo 113-8656, Japan.}
\affiliation{The Institute for Solid State Physics, University of Tokyo, Kashiwa, Chiba 277-8581, Japan.}
\author{Kazuya Miyagawa}
\affiliation{Department of Applied Physics, University of Tokyo, Tokyo 113-8656, Japan.}
\author{Svetlana A. Torunova}
\affiliation{Institute of Problems of Chemical Physics RAS, Chernogolovka 142432, Russia.}
\author{Natalia Drichko}
\affiliation{The Institute for Solid State Physics, University of Tokyo, Kashiwa, Chiba 277-8581, Japan.}
\affiliation{The Institute for Quantum Matter and the Department of Physics and Astronomy, The Johns Hopkins University, Baltimore, MD 21218, USA.}
\author{Elena I. Zhilyaeva}
\affiliation{Institute of Problems of Chemical Physics RAS, Chernogolovka 142432, Russia.}
\author{Kazushi Kanoda}
\email{kanoda@ap.t.u-tokyo.ac.jp}
\affiliation{Department of Applied Physics, University of Tokyo, Tokyo 113-8656, Japan.}
\affiliation{Max Planck Institute for Solid State Research,	Heisenbergstrasse 1, Stuttgart 70569, Germany}
\affiliation{Physics Institute, University of Stuttgart, Pfaffenwaldring 57, Stuttgart 70569, Germany}
\affiliation{Department of Advanced Materials Science, University of Tokyo,	Kashiwa 277-8561, Japan}

\date{\today}

\begin{abstract}
We investigated the local magnetism of the organic conductor, $\kappa$-(ET)$_{2}$Hg(SCN)$_{2}$Cl, with a quasi-triangular lattice of weakly dimerized molecules through $^{13}$C NMR spectroscopy. 
The NMR spectra and nuclear relaxation show that charge disproportionation occurs, associated with the metal-insulator transition at 31 K.
The relaxation rate indicates that the paramagnetic spins in the insulating phase undergo a transition into a spin-singlet ground state with the emergence of orphan spins, a possible valence-bond-glass state.
The present results are in high contrast to the spin-cluster paramagnetism of the electric dipole-liquid candidate, $\kappa$-(ET)$_{2}$Hg(SCN)$_{2}$Br, having nearly identical material parameters. This fact indicates that these two systems are on the verge between distinct phases in both charge and spin degrees of freedom; a spin-singlet charge-ordered state versus a spin-active Mott insulating state, competing with each other on a triangular lattice of dimerized sites.
\end{abstract}

\maketitle

\section{Introduction}
\begin{figure}
	\includegraphics{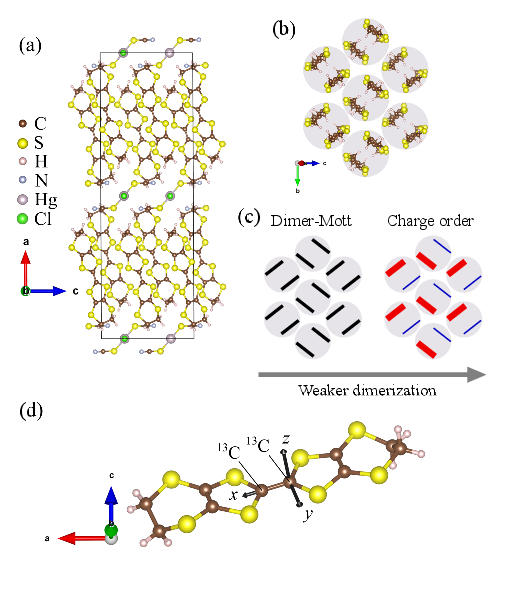}
	\caption{\label{fig:structure} Crystal structure. (a) Crystal structure of $\kappa$-Hg-Cl~\cite{PRB-2014-Drichko}. (b) $\kappa$-type ET arrangement in the conducting layer. (c) Schematic picture of the dimer-Mott and charge-ordered states. Solid lines represent ET molecules. Two ET molecules in a dimer are equivalent in the dimer-Mott state (left) and charge-disproportionated in the charge-ordered state (right). (d) ET molecule. The central double-bonded carbons were enriched by $^{13}$C isotope for $^{13}$C NMR experiments. The orthogonal axes colored black indicate the molecular principal-axes $x$, $y$ and $z$. (a), (b) and (d) were created using the visualizing software VESTA~\cite{VESTA}.}
\end{figure}

The family of layered organic conductors, $\kappa$-(ET)$_{2}$\textit{X}, which consists of electron donor, bis(ethylenedithio)tetrathiafulvalene, ET, and electron accepter, \textit{X}, has provided a rich variety of phenomena associated with strongly electron correlation, such as the metal-insulator transition and unconventional superconductivity~\cite{AnnRev-2011-Kanoda}.
An important feature for understanding the electric properties of $\kappa$-(ET)$_{2}$\textit{X} is found in the crystal structure (Fig.~\ref{fig:structure}); two ET molecules which face each other in an ET layer form a dimer.
When the ET dimerization is strong enough, the intradimer degrees of freedom become negligible and, as a consequence, the system can be regarded as an anisotropic triangular lattice consisting of ET dimers, with one hole per lattice site, which provides a suitable situation for studying the Mott physics on geometrically frustrated lattices~\cite{JPSJ-1996-Kino}.

Recently, the system with relatively weak ET dimerization, $\kappa$-(ET)$_{2}$Hg(SCN)$_{2}$\textit{Y} (\textit{Y} =Cl, Br), abbreviated as $\kappa$-Hg-\textit{Y} hereafter, has shed light on the charge degrees of freedom within an ET dimer, increasing diversity in electromagnetic properties emerging in $\kappa$-(ET)$_{2}$\textit{X}~\cite{PRB-2010-Hotta,PRB-2016-Naka,PRB-2020-Jacko}.
Due to small intradimer transfer energy in $\kappa$-Hg-\textit{Y}~\cite{PRL-2018-Gati,PRB-2020-Jacko}, the intermolecular Coulomb interaction within an ET dimer can drive the system into a charge-ordered insulator with intradimer charge disproportionation [Fig.~\ref{fig:structure}(c)].
Indeed, $\kappa$-Hg-Cl is suggested to exhibit a charge disproportionation below $T_{\textrm{CO}}$ of approximately 30 K by optical vibrational spectroscopy~\cite{PRB-2014-Drichko,PRB-2017-Ivek,Science-2018-Hassan} and the emergence of ferroelectricity~\cite{PRL-2018-Gati}.
Interestingly, however, recent Raman spectroscopy has suggested that this symmetry-broken charge-imbalanced state is not robust but gets into to a static or slowly fluctuating charge-disordered state at low temperatures below 15 K~\cite{npj-2020-Hassan}.
This charge-disordered state resembles the possible dipole-liquid state in $\kappa$-Hg-Br, in which the electric dipole arising from charge disproportionation within the ET dimer remains fluctuating down to low temperatures~\cite{Science-2018-Hassan}, and implies that $\kappa$-Hg-Cl is also situated near the boundary between the dimer-Mott and charge-ordered states.

The magnetic properties of $\kappa$-Hg-\textit{Y} have been also studied lately.
Especially, the dipole liquid material $\kappa$-Hg-Br shows a soft magnetic response, which is explained well by considering the coupling of spin and charge degrees of freedom in $\kappa$-Hg-Br~\cite{npj-2021-Yamashita}. 
To date, several experimental studies reported the magnetic properties of $\kappa$-Hg-Cl~\cite{PRL-2018-Gati,PhysicaB-2012-Yasin,PRB-2022-Drichko,PRB-2020-Pustogow}.
Magnetic susceptibility decreases upon cooling below 25 K, suggesting a spin singlet formation, but shows a Curie-like upturn below 15 K~\cite{PhysicaB-2012-Yasin,PRB-2022-Drichko}, consistent with the $^{1}$H NMR observation of slowing down of magnetic fluctuations typical of the impurity-dominant relaxation~\cite{PRB-2020-Pustogow}.
The origin of this impurity-like behavior was proposed to be paramagnetic spins located at charge-order domain walls or impurity spins~\cite{PRB-2022-Drichko,PRB-2020-Pustogow}, which makes it difficult to observe the bulk magnetic properties in $\kappa$-Hg-Cl.

In the present work, we investigated the spin states of $\kappa$-Hg-Cl by conducting $^{13}$C NMR measurements of a single crystal of $\kappa$-Hg-Cl,  in which the central double-bonded carbons in ET were enriched by $^{13}$C isotope [Fig.~\ref{fig:structure}(d)].
$^{13}$C NMR is superior to $^{1}$H NMR for studying the local static and dynamical spin susceptibilities of $\pi$ electrons due to the large hyperfine coupling constant of $^{13}$C nuclei.
As a result, we observed a spectral broadening due to a charge disproportionation below 31 K, followed by spectral narrowing accompanied with a sharp decrease in the $^{13}$C nuclear spin-lattice relaxation rate, $1/T_{1}$, below approximately 25 K, as an indication of the spin-singlet ground state.
The $^{13}$C NMR spectrum has a multi-peak structure at the lowest measured temperature of 4 K, indicative of inhomogeneous charge disproportionation within an ET dimer, possibly related to the partial melting of charge order suggested by the Raman spectroscopy~\cite{npj-2020-Hassan}.

\section{Experimental Methods}
Single crystals of $\kappa$-Hg-Cl and $\kappa$-Hg-Br were synthesized by the electrochemical method.
We conducted $^{13}$C NMR experiments using single crystals of $\kappa$-Hg-Cl, in which the central double-bonded carbons in ET were enriched by $^{13}$C isotope.
The magnetic field of 6.00 T was applied nearly parallel to the $a$ axis. $^{13}$C NMR spectra and nuclear spin-lattice relaxation time $T_{1}$ were obtained by the standard solid-echo method using the pulse sequence $t_{\pi/2}$-$\tau$-$t_{\pi/2}$ with the $\pi/2$ pulse width $t_{\pi/2}$ and the interval time $\tau$.
Throughout the measurements, $\tau$ was 30 $\mu$s, and $t_{\pi/2}$ was in the range of 2--3 $\mu$s.
For comparison, we also investigated $^{13}$C NMR spectra and $1/T_{1}$ of $\kappa$-Hg-Br in a field of 6.00 T, using a single crystal.
Because the spectra at temperatures below 30 K for $\kappa$-Hg-Br are too wide to be covered by a rf pulse, we measured spectra at sequential frequencies at intervals of 50 or 100 kHz and summed them to obtain the full spectrum.
The results for $\kappa$-Hg-Br were in good agreement with the previous report~\cite{PRB-2020-Le}.

\section{Results}
\subsection{$^{13}$C NMR spectra}
\begin{figure*}
	\includegraphics{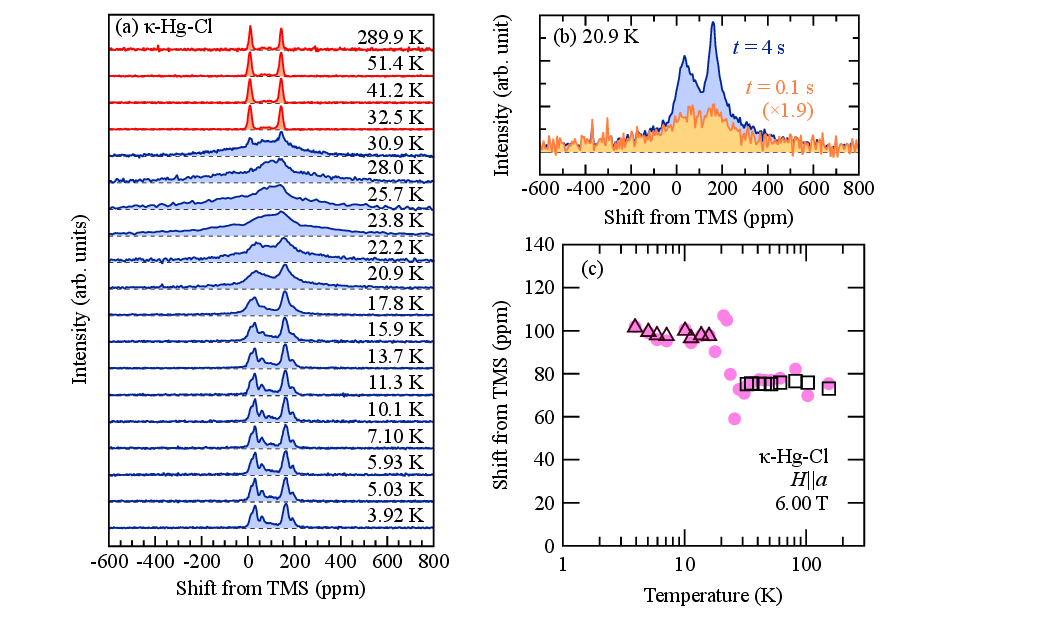}
	\caption{\label{fig:spectra} $^{13}$C NMR spectra of $\kappa$-Hg-Cl in the magnetic field of 6.00 T nearly parallel to the $a$ axis. (a) Temperature dependence of $^{13}$C NMR spectra. (b) $^{13}$C NMR spectra at 20.9 K obtained at times $t =$4 and 0.1 s after the nuclear magnetization saturation. The spectral intensity for $t =$0.1 s was multiplied by 1.9, which was derived from $[1-\exp(-4/T_{1,\textrm{short}})]/[1-\exp(-0.1/T_{1,\textrm{short}})]$ with $1/T_{1,\textrm{short}} =7.5$ s$^{-1}$, giving the fast-component spectra expected at $t = 4$ s. (c) Temperature dependence of the spectral shift of $\kappa$-Hg-Cl, determined by the first moment of the entire spectrum (closed circles), the average of the center positions obtained by the two-gaussian fit of the entire spectrum (open squares), and the average of the first moments calculated for the left- and right-side spectra in the shift ranges from $-100$ to $80$ ppm and from 120 to 300 ppm, respectively (open triangles).}
\end{figure*}

Figure~\ref{fig:spectra} shows the temperature dependence of $^{13}$C NMR spectrum of $\kappa$-Hg-Cl.
At room temperature, the spectrum consists of the Pake doublet, namely, two spectral lines split by the nuclear dipole interaction between neighboring $^{13}$C nuclei in ET.
The splitting width of the Pake doublet is determined by the distance between the neighboring $^{13}$C nuclei, $r$, and the angle between the direction of the magnetic field and the $^{13}$C-$^{13}$C vector, $\eta$, through $3\gamma_{n}^2\hbar(1-3\cos^2{\eta})/2r^{3}$ with the $^{13}$C nuclear gyromagnetic ratio $\gamma_{n}$ and reduced Planck constant $\hbar$. 
The observed line splitting width was 8.6 kHz, which was larger than the theoretical value 7.85 kHz which was obtained from $r = 1.36$ {\AA} and $\eta = 17.26^\circ$~\cite{PRB-2014-Drichko} for the case in which the external field was applied parallel to $a$-axis.
Assuming that the discrepancy between the experimental and theoretical values is entirely attributable to crystal misalignment, a misalignment of no more than 7 degrees can explain for the difference. In reality, however, there are additional uncertainties beyond misalignment. For example, the distance between the $^{13}$C nuclei can differ from 1.36 {\AA}, since this value was obtained from X-ray diffraction measurements, which probe the electron density, whereas the width of the Pake doublet is determined by the relative positions of the nuclei rather than the electrons.

On cooling, the spectrum of $\kappa$-Hg-Cl remains unchanged in the metallic phase down to $T_{\textrm{CO}}$ of $\sim$31 K.
Invariance of the spectral position with temperature variation is consistent with the uniform magnetic susceptibility being nearly constant in the metallic state~\cite{PRB-2022-Drichko}.
Upon further cooling, the spectrum of $\kappa$-Hg-Cl abruptly broadens below $T_{\textrm{CO}}$, indicative of an increase in the distribution of local spin susceptibility.
Except for the sharp lines in the spectrum at 30.9 K, which is  attributed to residual metallic domains in the coexistence region of the first-order transition, the main broad spectrum just below $T_{\textrm{CO}}$ consists of two components with different relaxation times; the broad and narrow spectra with the short and long relaxation times can be assigned to signals from the charge-rich and charge-poor ET sites, respectively [Fig~\ref{fig:spectra}(b)].
The broad spectrum for the charge-rich sites is probably a broadened quartet which is observed when the local fields on the neighboring $^{13}$C nuclei are nonequivalent such as in other charge-ordered systems~\cite{PRB-2008-Chiba,JPSJ-2019-Miyagawa}. The absence of clear peak structure indicates intrinsic or extrinsic inhomogeneities in local fields.
The spectrum remains nearly unchanged down to 24 K and exhibits rapid narrowing upon further cooling.
The low-temperature spectra below 7 K are nearly temperature-independent, indicating the nonmagnetic nature of $\kappa$-Hg-Cl.
At these low temperatures, the spin shift is expected to be negligibly small, and the spectral position is solely determined by the chemical shift.
The slightly broadened and multi-peaked spectral shape suggests the distributed charge density, because the chemical shift for ET-based compounds is dependent on the valence of ET~\cite{JPSJ-2009-Kawai,PRB-2011-Hirata} (see the Appendix~\ref{appendix:lowTspectrum} for a detailed analysis).

Figure~\ref{fig:spectra}(c) shows the temperature dependence of the spectral shift of $\kappa$-Hg-Cl.
The spectral shift is characterized by the first moment of the entire spectrum, plotted as closed circles in Fig.~\ref{fig:spectra}(c) in the entire temperature range; we also evaluated the spectral shift by the average of the center positions of two-Gaussian fit of the spectra in the metallic state, and by the average of first moments of the left- and right-side spectra in the range from $-100$ to 80 ppm and from 120 to 300 ppm, respectively at the temperatures below 16 K, each of which gave a close value to the first moment of the entire spectrum.
The spectral shift is nearly constant in the high-temperature metallic phase above $T_{\textrm{CO}}$, as is expected from the Pauli-paramagnetic behavior~\cite{PRB-2022-Drichko}.
Upon cooling below approximately 25 K, the spectral shift increases down to approximately 15 K.
The onset of the increase in the spectral shift, 25 K, coincides with the onset of the decrease in the magnetic susceptibility~\cite{PRB-2022-Drichko}, consistent with a spin-singlet formation.
The saturated value of the shift at low temperatures, $\sim100$ ppm, corresponds to the average of chemical shifts of spectral lines consisting of the spectrum.
Given the value 100 ppm for the chemical shift, the hyperfine coupling constant in the metallic phase was evaluated to be $-0.70$ kOe/$\mu_{\textrm{B}}$ from the Knight shift, $-25$ ppm, and spin susceptibility, 0.0004 emu/mol f.u.~\cite{PRB-2022-Drichko}.
The negative hyperfine-coupling constant in the present field configuration is consistent with the anisotropic hyperfine tensor~\cite{PRB-1995-Kawamoto,PRB-2003-Smith}.
No remarkable change of the spectral shift toward the lowest temperature indicates that the Cuire-like increase in the susceptibility~\cite{PRB-2022-Drichko} does not originate from the local spin susceptibility in bulk.
Therefore, we concluded that the low-temperature behavior of magnetic susceptibility below 10 K is determined by the contribution from a minority of spins, very likely orphan spins out of the spin singlet background.
 
\subsection{$^{13}$C nuclear spin-lattice relaxation rate}
\begin{figure}
	\includegraphics{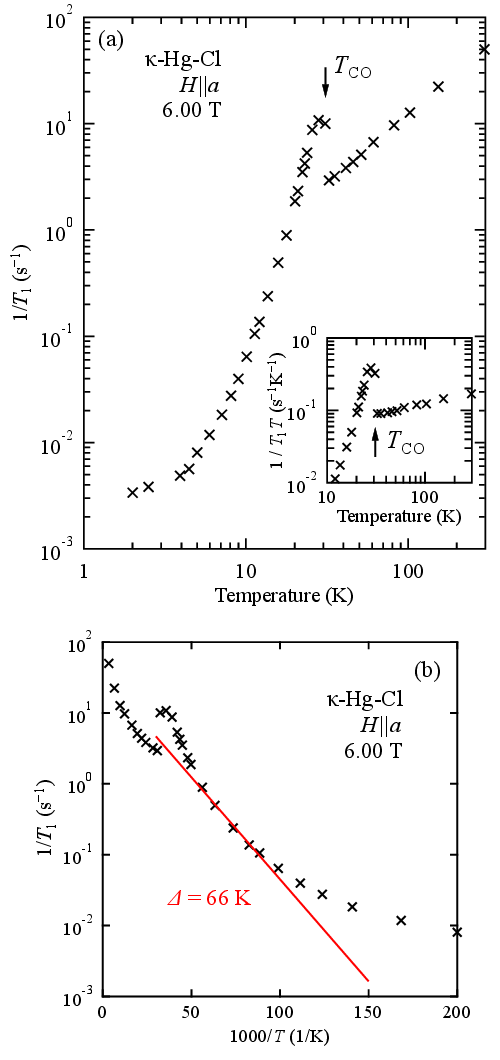}
	\caption{\label{fig:T1} $^{13}$C nuclear spin-lattice relaxation rate, $1/T_{1}$. (a) Temperature dependence of $1/T_{1}$ determined by the stretched exponential fits to the $^{13}$C nuclear magnetization recovery curve. The inset plots $1/T_{1}T$ versus temperature in the 10--300 K range. (b) The plot of $^{13}$C $1/T_{1}$ against the inverse of temperature.}
\end{figure}

Figure~\ref{fig:T1}(a) shows the temperature dependence of the $^{13}$C nuclear spin-lattice relaxation rate, $1/T_{1}$, obtained by the stretched exponential fit to the nuclear magnetization recovery curve, $1-M(t)/M(\infty)$, where $M(t)$ is the nuclear magnetization at time $t$ after saturation and $M(\infty)$ is the saturation value, with the expression of $\exp[-(t/T_{1})^\beta]$ with parameters $T_{1}$ and $\beta$ (see Appendix~\ref{appendix:relaxation} for the recovery curves).
The exponent $\beta$, which measures the degree of the inhomogeneity in the relaxation of $^{13}$C nuclear magnetization, remains close to 1 in the metallic state above $T_{\textrm{CO}}$, indicating a homogeneous relaxation.
Below $T_{\textrm{CO}}$, $\beta$ shows a temperature-dependent variation suggesting a change in the relaxation mechanism, as discussed later.
As seen in the inset of Fig.~\ref{fig:T1}(a), in the metallic phase above $T_{\textrm{CO}}$, $1/T_{1}$ is roughly proportional to temperature, $T$, especially below 100 K, consistent with the previous $^{1}$H NMR study~\cite{PRB-2020-Pustogow}.
From the $1/T_{1}T$ value of 0.1 s$^{-1}$K$^{-1}$ and Knight shift, $K$, of $-25$ ppm, we estimated the enhancement factor, $K(\alpha)$, in the Korringa relation, $(1/T_{1}T) K^{-2} (\hbar/4\pi k_{\textrm{B}})( \gamma_{e}/\gamma_{n}) = \zeta(\theta,\varphi) K(\alpha)$, to be $\sim10$,
where $\gamma_{e}$ is the gyromagnetic ratio of an electron and $\zeta(\theta,\varphi)$ is the geometrical factor coming from the anisotropy of the hyperfine coupling tensor and is of the order of $10^{1}$ in the present magnetic-field geometry (see Appendix~\ref{appendix:Korringa} for details).
The $K(\alpha)$ value of $\sim$10 is comparable to the values in $\kappa$-(ET)$_{2}$Cu(NCS)$_{2}$, $\kappa$-(ET)$_{2}$Cu[N(CN)$_{2}$]Br, and deuterated $\kappa$-(ET)$_{2}$Cu[N(CN)$_{2}$]Br~\cite{PRB-1995-Soto,ChemRev-2004-Miyagawa,PRR-2023-Furukawa}, which are half-filled-band metals situated near the Mott transition boundary, and suggests strong antiferromagnetic correlation in the metallic state in $\kappa$-Hg-Cl located near the metal-insulator phase boundary.

\begin{figure}
	\includegraphics{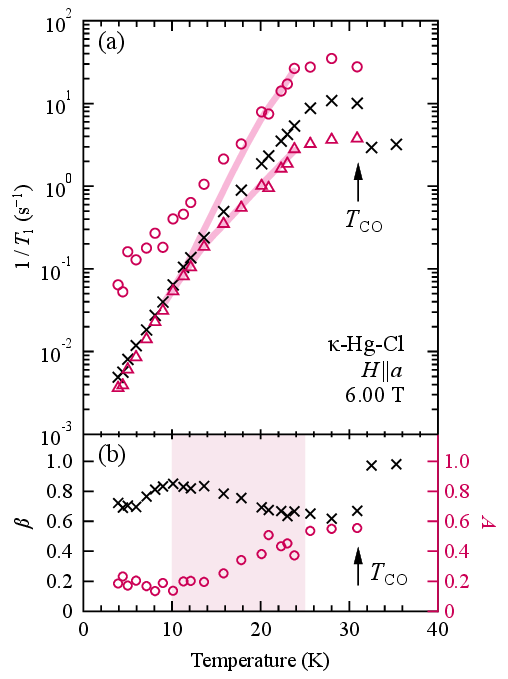}
	\caption{\label{fig:T1_CO} Analyses of $^{13}$C nuclear spin-lattice relaxation. (a) $^{13}$C nuclear spin-lattice relaxation rate, $1/T_{1}$, determined by the stretched exponential function, $\exp[-(t/T_{1})^\beta]$, (crosses) and the double exponential function,  $A\exp(-t/T_{\textrm{1,short}}) + (1-A)\exp(-t/T_{\textrm{1,long}})$, (circles and triangles) to the $^{13}$C nuclear magnetization recovery curve. The bold curves depict the expected behaviors of $1/T_{1}$ at the charge-rich and charge-poor sites (see text). (b) Temperature dependence of the fitted values of $\beta$ (crosses, left axis) and $A$ (circles, right axis). }
\end{figure}

Upon cooling, $1/T_{1}$ shows an abrupt increase just below $T_{\textrm{CO}}$, reflecting an increase in the dynamic spin susceptibility.
The recovery curve consists of two components with different relaxation time, $T_{\textrm{1,short}}$ and $T_{\textrm{1,long}}$, which correspond to the relaxation times at the charge-rich and charge-poor sites, respectively [see also Fig.~\ref{fig:spectra}(b)].
The double-exponential fit to the recovery curve with the form of $A\exp(-t/T_{\textrm{1,short}}) + (1-A)\exp(-t/T_{\textrm{1,long}})$ with the fraction of short-$T_{1}$ component, $A$, yielded the temperature-insensitive values, $A \sim0.55\pm0.01$, $1/T_{\textrm{1,short}} \sim30\pm4$ s$^{-1}$ and $1/T_{\textrm{1,long}} \sim3.6\pm0.3$ s$^{-1}$ in the temperature range of 25--31 K (Fig.~\ref{fig:T1_CO}).
The $A$ value close to 0.5 suggests that volume fractions of the charge-poor and charge-rich sites are nearly equal, and the temperature-insensitive behavior of $1/T_{1}$ is typical for localized spins.
The ratio of square root of $1/T_{1}$, $[(1/T_{\textrm{1,short}})/(1/T_{\textrm{1,long}})]^{1/2}$ of $\sim3$ is larger than the charge disproportionation ratio, 1.5, determined from the optical vibrational spectroscopy~\cite{PRB-2014-Drichko,Science-2018-Hassan}.
This is reasonably explained in terms of the the so-called form-factor effect in $1/T_{1}$ in the presence of enhanced antiferromagnetic fluctuations; namely, at the charge-poor sites, local fields from the neighboring antiferromagnetically aligned spins tend to be canceled so that $1/T_{1}$ at those sites are suppressed, leading to a larger value of the $1/T_{1}$ ratio than expected from the charge-density ratio.
Below 25 K, $1/T_{1}$ rapidly decreases with temperature, which indicates a spin singlet formation. 
Using the $1/T_{1}$ values in the narrow temperature range 11--18 K, the spin excitation gap was evaluated to be $\Delta = 66$ K through $1/T_{1} \sim \exp(-\Delta/T)$.
However, this value should be taken as a nominal value because the temperature dependence of $1/T_{1}$ is not well characterized by a single energy gap as shown in Fig.~\ref{fig:T1}(b). This non-activation behavior may be due to the spin diffusion from the impurity site.
The obtained value is smaller than the reported charge gap of 140 K, observed in the optical conductivity~\cite{PRB-2014-Drichko}, or 432 K, evaluated by the Raman scattering intensity~\cite{npj-2020-Hassan}, which ensures that the energy gap observed in $1/T_{1}$ is that for a spin singlet-triplet excitation.

 Simultaneously with the decrease of $1/T_{1}$, the recovery curve of the nuclear magnetization gradually changes from the double exponential behavior with $A = 0.5$ to the stretched exponential function with an increase in the $\beta$ value on cooling.
This apparent recovery to homogeneous relaxation is similar to the behavior of another charge ordering system, $\theta$-(ET)$_{2}$RnZn(SCN)$_{4}$ ($\theta$-RbZn)~\cite{PRB-2000-Miyagawa}; spin-spin relaxation ($T_{2}$ process) between $^{13}$C nuclei, which averages out the inhomogeneous $1/T_{1}$ values at the neighboring charge-rich and charge-poor sites, becomes more effective as  $T_{1}$ becomes longer at lower temperatures.
Similarly, the $1/T_{1}$ values for the charge-rich and charge-poor sites are expected to asymptotically approach each other with decreasing temperature. 
Indeed, the two-component fits to the relaxation curves yield that their fraction gradually deviates from 0.5:0.5 at 25K to 0.2:0.8 at 10 K on cooling, which suggests that the two components no longer reflect the charge-rich and charge-poor sites.
It is likely that the $1/T_{1}$ values of the two sites gradually merge on cooling as depicted by the conjectured lines in Fig.~\ref{fig:T1_CO}(a) to jointly give the major component with the longer $T_{1}$ and the other short $T_{1}$ component shows up as another relaxation component.
Considering that the magnetic susceptibility exhibits an upturn below the order of 10K~\cite{PRB-2022-Drichko}, the origin of the short $T_{1}$ component is attributable to orphan spins.
Note that the NMR experiments were conducted in a magnetic field of 6 Tesla.
In this field strength, the free spins become increasingly polarized as the temperature decreases, starting from above 10 K. Consequently, the fluctuations of the free spins are progressively suppressed with decreasing temperature, which can lead to a gradual suppression of $1/T_{1}$ of the first component at lower temperatures instead of the conventional temperature-insensitive behavior.
It is noted that the intensity fraction, 20\%, of the fast relaxation component does not mean the fraction of impurity sites but the fraction of sites at which nuclear relation is enhanced by the nuclear spin diffusion to the impurity sites~\cite{PR-1960-Blumberg}.
Indeed, the ``impurity'' spin density evaluated from the Curie ﬁt to magetic susceptibility was reported to be 1.6\% at most~\cite{PRB-2020-Pustogow}.
Moreover, at low temperatures below 10 K and under the magnetic field of 6T, the nuclear spins on the impurity sites suffer from large internal magnetic fields of the order of kOe due to the hyperfine interaction with electron spins and be out of the observed spectrum.
Such nuclear spin relaxation dominated by impurity-like moments at low temperatures has already been suggested by the previous $^{1}$H NMR study~\cite{PRB-2020-Pustogow}. We note that, in the charge ordered state, the temperature dependence of $^{13}$C $1/T_{1}$ is distinct from that of $^{1}$H $1/T_{1}$,  in contrast to the similar behavior between the two with the temperature-insensitive ratio of $^{13}$C and $^{1}$H $1/T_{1}$'s, $\sim$120, in the high-temperature metallic phase, which is determined solely by the different values of nuclear gyromagnetic ratio and hyperfine coupling constatnt for $^{13}$C and $^{1}$H nuclei in ET.
Regarding the different temperature dependences of $^{13}$C and $^{1}$H $1/T_{1}$'s, the exponent $\beta$ in the stretched exponential fit of the nuclear relaxation just below  $T_{\textrm{CO}}$ was $\beta \sim 0.9$ for $^{1}$H NMR~\cite{PRB-2020-Pustogow} and $\beta \sim 0.6$ for $^{13}$C NMR.
This indicates the $T_{2}$ process is more effective in $^{1}$H nuclear spin relaxation; the fast nuclear relaxation at the impurity site becomes more influential to surrounding sites through the spin-spin relaxation between neighboring nuclei and, as a result, the temperature dependence of $1/T_{1}$ for minor sites was more emphasized in $^{1}$H $1/T_{1}$ than in $^{13}$C $1/T_{1}$~\cite{PR-1960-Blumberg}.
Another approach to extract $1/T_{1}$ of electronic systems exhibiting inhomogeneous relaxation is based on the inverse Laplace transform (ILT)~\cite{PRB-2020-Singer,NatPhys-2021-Wang}. 
We applied the ILT method to analyze the $T_{1}$ relaxation of $^{13}$C nuclear magnetization. [See also Appendix~\ref{appendix:relaxation} and Figs.~\ref{fig:relaxation_ILT}(e)--(k). Figures~\ref{fig:relaxation_ILT}(g)--(k) display the $1/T_{1}$ profiles obtained at several temperatures through the ILT analysis. For comparison, we also present the ILT analysis for the non-magnetic charge-ordered system $\theta$-RbZn in Figs.~\ref{fig:relaxation_ILT}(e) and (f).]
At a temperature of 25.6 K, where the charge-ordered state is paramagnetic, the two peaks in the $1/T_{1}$ distribution correspond to the $1/T_{1}$ values from fitting the sum of two exponential functions. Below 24 K, where spins start to form singlets, multiple peaks appear. The $1/T_{1}$ value of the prominent peak at a lower frequency corresponds to the value of the spin-singlet phase deduced from the two-exponential fit. 
At 20.1 K, just below the singlet transition, the additional two peaks that emerge at frequencies similar to those at 25.6 K can reasonably be attributed to the coexisting paramagnetic phase. At lower temperatures, where spin singlets are well-developed, the additional peaks contribute minimally and may be related to relaxation by orphan spins. 
We are unsure why two additional peaks appear at 15.8 K and 3.92 K and only a single additional peak is observed at 10.1 K. The two peaks may be related to charge-rich and charge-poor sites with staggered moments of different magnitudes, induced by nearby impurity spins; the lower-frequency peak at 10.1 K may overlap with the main peak. These peak profiles are sensitive to the parameter choices in the analysis, and therefore do not warrant further discussion. 

\section{Discussion}\label{Sec_Discussion}
\begin{figure*}
	\includegraphics{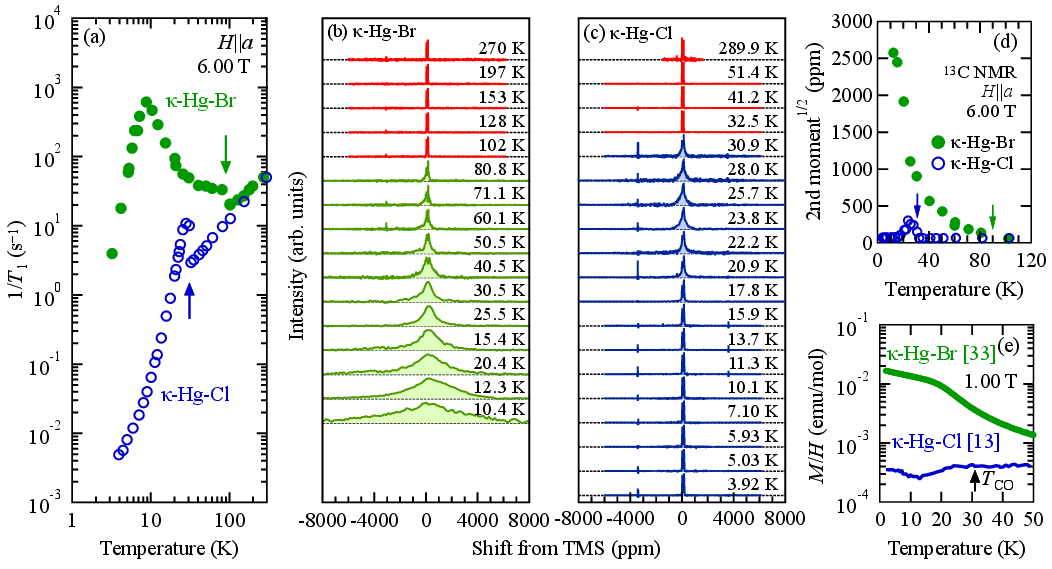}
	\caption{\label{fig:all_k-Hg-Y} $^{13}$C NMR results and magnetic susceptibilities for $\kappa$-Hg-Cl and $\kappa$-Hg-Br. (a) Temperature dependence of $^{13}$C $1/T_{1}$. The arrows indicate the metal-insulator transitions for $\kappa$-Hg-Cl and $\kappa$-Hg-Br, $T_{\textrm{CO}}$ and $T_{\textrm{MI}}$, respectively. (b,c) $^{13}$C NMR spectra of (b) $\kappa$-Hg-Br and (c) $\kappa$-Hg-Cl. [The data shown in (c) is reproduced from the data in Fig.~\ref{fig:spectra}(a).] (d) Temperature dependence of the square root of second moment of $^{13}$C NMR spectra. (e) Magnetic susceptibilities measured at a magnetic field of 1.00 T taken from Refs.~\cite{PRB-2022-Drichko,SciAdv-2022-Urai}.}
\end{figure*}
\subsection{Spin singlet formation in $\kappa$-Hg-Cl}
As described above, the $^{13}$C NMR characteristics of the insulating state in $\kappa$-Hg-Cl is well explained by charge disproportionation below 31 K and spin singlet formation below 25 K.
At low temperatures below approximately 10 K, additional inhomogeneity shows up; the spectra indicates that the charge density has a nonuniform distribution. 
The indication of spin-singlet formation is also consistent with the vanishingly small electronic specific heat coefficient $\gamma$ for $\kappa$-Hg-Cl~\cite{Science-2018-Hassan}.
The inhomogeneous charge distribution captured by $^{13}$C NMR spectra at low temperatures, at which the magnetic hyperfine field is negligibly small and the chemical shift determies the spectral profile, accord with the coexistence of ET molecules with different valences~\cite{npj-2020-Hassan}.

Now we discuss the possible mechanism of the spin singlet formation.
The present results are reminiscent of  $\theta$-RnZn, which exhibits charge disproportionation below the metal-insulator transition temperature, $\sim$190 K~\cite{PRB-2000-Miyagawa,PRL-2004-Chiba,PRB-2002-Yamamoto,PhysicaB-2009-Drichko}, and  shows spin-gapped behavior in magnetic susceptibility below 30 K~\cite{PRB-1998-Mori}. 
For  $\theta$-RnZn, the x-ray diffraction study reported the lattice symmetry breaking below 20 K and suggested a spin-Peierls mechanism for the nonmagnetic transition~\cite{JPSJ-2007-Watanabe}.
For $\kappa$-(ET)$_{2}$\textit{X}, geometrical frustration can drive spins into the valence bond solid (VBS) state concomitant with a symmetry breaking, in the presence of a strong electron-phonon coupling~\cite{PRB-2011-Dayal}.
Although the results of x-ray diffraction measurements of $\kappa$-Hg-Cl reported so far did not find change of the lattice symmetry from the $C2/c$ space group down to 10 K~\cite{PRB-2014-Drichko}, thermodynamic studies detected lattice response on metal-insulator transition at $T_{\textrm{CO}}$~\cite{PRL-2018-Gati}.
Further investigation to clarify the presence or absence of lattice symmetry breaking is warranted.
Considering the broadened charge-density distribution in the charge-ordered state and the triangular lattice which causes spin frustration, the valence bond glass (VBG)~\cite{EPL-2008-Tarzia,PRL-2010-Singh} is among the conceivable states realized in $\kappa$-Hg-Cl at the low temperatures.
Riedl \textit{et al.} suggested a possible VBG state in $\kappa$-(ET)$_{2}$\textit{X}~\cite{NCommun-2019-Riedl}; they studies the magnetism of the dimer-Mott state, in which spins-1/2 are localized on the triangular lattice of ET dimers, and suggested the formation of VBG-like randomly configured singlets with a number of spin defects.
Although $\kappa$-Hg-Cl, a charge-disproportionated insulator, is different from the dimer Mott insulator postulated by their model, the resultant states of the two quite resemble each other; singlet formation, no long-range order and a sizable number of spin defects.
We note that the density of the Curie-like spins in $\kappa$-Hg-Cl, 1.6\%, is too large to be considered as extrinsic impurity spins in organic conductors~\cite{NPhys-2015-Yoshida}; it appears rather reasonable to regard it as an emerge of magnetic defects upon the singlet formation. 
As mentioned in the Introduction,  $\kappa$-Hg-Cl exhibits dual frustration in both charge and spin degrees of freedom. With regard to charge frustration, the system lies near the boundary between a Mott insulator and a charge-ordered insulator. This positioning, as well as the triangular lattice geometry of ET dimers, can lead to the emergence of inhomogeneous charge disproportionation and/or charge defects, which are likely to produce a significant number of orphan spins when the system transitions to a singlet ground state.
Although this issue requires further consideration, the orphan spins generated in this system may be related to such frustration effects. 

\subsection{Contrasting spin states in $\kappa$-Hg-Cl and $\kappa$-Hg-Br}
Remarkably, the $^{13}$C $1/T_{1}$ behavior in the insulating state of $\kappa$-Hg-Cl is qualitatively distinct from that of $\kappa$-Hg-Br, which shows an increase in $1/T_{1}$ below the metal-insulator transition and a pronounce peak at approximately 8 K [Fig.~\ref{fig:all_k-Hg-Y}(a)], although these two isostructural compounds have similar transfer integrals~\cite{PRB-2020-Jacko}; indeed, the values of $1/T_{1}$ in the high-temperature metal phases are close to each other.

In Ref.~\cite{SciAdv-2022-Urai}, the $1/T_{1}$ behavior in the insulating phase of $\kappa$-Hg-Br is discussed in terms of the vital fluctuations of spin clusters that slow down upon cooling without ordering or freezing. 
The distinct spin states of the two systems at low temperatures are also clear in the spectra.
Figures~\ref{fig:all_k-Hg-Y}(b) and (c) compare the $^{13}$C NMR spectra of the two systems acquired at a magnetic ﬁeld of 6.00 T.
Both compounds show spectral broadening just below the metal-insulator transition temperatures of $\sim$90 K and $\sim$31 K for $\kappa$-Hg-Br and $\kappa$-Hg-Cl, respectively.
However, its temperature evolution upon further cooling is contrasting between the two.
Whereas the spectrum of $\kappa$-Hg-Cl turns to a narrowing below 24 K due to the spin-singlet formation, the spectrum of $\kappa$-Hg-Br keeps broadening down to the lowest temperature, at which the line width becomes 30 times larger than that in $\kappa$-Hg-Cl, as seen in Fig.~\ref{fig:all_k-Hg-Y}(d), indicating increasing inhomogeneous internal fields in $\kappa$-Hg-Br contrary to the vanishing local fields in $\kappa$-Hg-Cl. 
Figure~\ref{fig:all_k-Hg-Y}(e) shows the spin susceptibilities of the two systems, which are also contrasting; the susceptibility in $\kappa$-Hg-Br shows a prominent increase with decreasing temperature~\cite{SciAdv-2022-Urai,PRB-2018-Hemmida,npj-2021-Yamashita}, reaching a value nearly two orders of magnitude larger than that in $\kappa$-Hg-Cl.
As discussed in detail in Ref.~\cite{SciAdv-2022-Urai}, the comparative analyses of the NMR linewidth and uniform spin susceptibility for $\kappa$-Hg-Br suggest the emergence of an anomalous spin state with field-induced antiferromagnetic and ferromagnetic moments unlike the spin-singlet state in $\kappa$-Hg-Cl.
The contrasting spin states in $\kappa$-Hg-Br and $\kappa$-Hg-Cl, clearly demonstrated by NMR and magnetic susceptibility suggests that strongly correlated electronic systems with triangular lattices of weakly dimerized molecules have highly competing spin states as well as charge states and $\kappa$-Hg-Cl  and $\kappa$-Hg-Br are marginal states in the verge of the instability.

Finally, we point out a common aspect to the two systems in their paramagnetic states just below the metal-insulator transitions.
For $\kappa$-Hg-Cl, the spectral width determined by the square root of second moment of the spectrum [Fig.~\ref{fig:all_k-Hg-Y}(d)] is 150  ppm at 30.9 K, just below $T_{\textrm{CO}}$.
This value corresponds to $1.0\times10^{-2} \mu_{\textrm{B}}$/ET, assuming the $^{13}$C hyperﬁne coupling tensor obtained for the ET-based metal $\theta$-(ET)$_{2}$I$_{3}$~\cite{PRB-2012-Hirata}.
This value is five times larger than the magnetic moment estimated from the susceptibility, 0.0004 emu/mol f.u. [Fig.~\ref{fig:all_k-Hg-Y}(e)], $2 \times 10^{-3} \mu_{\textrm{B}}$/ET at 6.00 T.
This means that the local moment probes captured by NMR have staggered configuration, similarly to the observation in $\kappa$-Hg-Br~\cite{SciAdv-2022-Urai}. 

\subsection{Estimation of Charge disproportionation from $^{13}$C NMR spectra}\label{subSec_SPC}
\begin{figure}
	\includegraphics{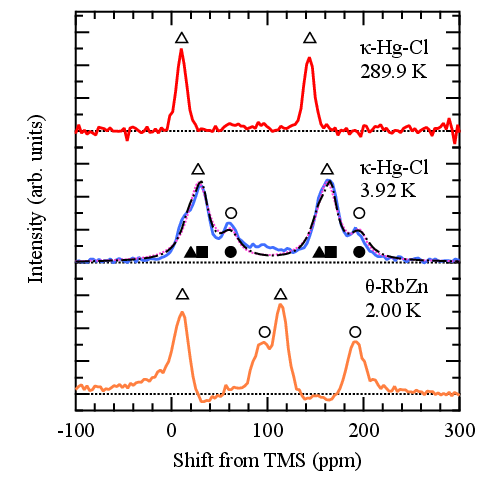}
	\caption{\label{fig:lowT_spectrum} $^{13}$C NMR spectra of $\kappa$-Hg-Cl at 289.9 K (top) and 3.92 K (middle) in 6.00 Tesla, and of $\theta$-RbZn at 2.00 K in 8.04 Tesla~\cite{PRB-2000-Miyagawa} (bottom). In each panel, pairs of spectral lines forming Pake doublets are indicated by the same symbols. In the middle panel, the broken and dash-dotted curves represent fits to four- and six-Lorentzian functions, corresponding to two and three pairs of Pake doublets, respectively; the resulting peak positions are indicated by the open and closed symbols (triangles, circles, and squares).}
\end{figure}

The low-temperature $^{13}$C NMR spectrum of $\kappa$-Hg-Cl is compared to that of a similar non-magnetic charge-ordered system, $\theta$-RbZn(SCN)$_{4}$, as shown in Fig.~\ref{fig:lowT_spectrum}.
The latter system exhibits two sets of Pake doublets, indicating the presence of two distinct sites arising from charge-rich and charge-poor molecules (0.85e$^{+}$ : 0.15e$^{+}$). The spectrum of  $\kappa$-Hg-Cl displays a similar feature, albeit with a smaller charge difference (0.64e$^{+}$ : 0.36e$^{+}$) (see Appendix~\ref{appendix:lowTspectrum} for derivation of the charge from the observed chemical shift). This is roughly represented by two sets of Pake doublets. However, the spectrum may also suggest fine structure, indicating a more complex charge disproportionation. For reference, a fit using three sets of Pake doublets is included in Fig. 6, which results in a charge disproportionation of 0.70e$^{+}$ : 0.45e$^{+}$ : 0.35e$^{+}$ with corresponding intensities of 1.7 : 1 : 2, consistent with the Raman profile~\cite{npj-2020-Hassan}. 
In $\kappa$-type ET arrangements, the two double-bonded$^{13}$C-enriched carbon sites are typically non-equivalent, potentially doubling the number of observable NMR lines. However, the spectrum of $\kappa$-Hg-Cl at ambient temperature (in the absence of charge order) reveals only a single Pake doublet, indicating that any non-equivalence is negligible in the spectral profile.

\section{Conclusions}
We investigated with $^{13}$C NMR the electronic state of the organic conductor, $\kappa$-Hg-Cl, with a quasi-triangular lattice of weakly dimerized ET molecules, which is isostructural to $\kappa$-Hg-Br suggested as a quantum electric-dipole liquid and showing anomalous magnetic properties.
The NMR spectra and relaxation rate demonstrate that the metal-insulator transition at 31 K is accompanied by a charge order with broadened charge disproportionation and followed by spin singlet formation below 25 K unlike the anomalously enhanced spin fluctuations in $\kappa$-Hg-Br.
The low-temperature nuclear relaxation profile in $\kappa$-Hg-Cl indicates the emergence of orphan spins in the singlet sea, which is reminiscent of a valence-bond-glass in conjunction with the absence of long-range structural order.
The contrasting electronic and magnetic properties of $\kappa$-Hg-Cl and $\kappa$-Hg-Br with quite similar lattice geometries and band structures provide evidence that strongly correlated electrons on triangular lattices can host a variety of charge and spin states.

\begin{acknowledgments}
	The authors are grateful to the late R. N. Lyubovskaya for synthesizing high-quality single crystals.
	The preparation of single crystals was supported by the Ministry of Science and Higher Education of the Russian Federation (Registration number 124013100858-3).
	This work was supported in part by JSPS Grants in Aid for Scientific Research (Grants Nos. JP18H05225, 20K20894, 20KK0060, 21K18144, 23K25815, 24K17004 and 24K21525).
	N.D. is grateful for the support of the Visiting Researcher's Program of the Institute for Solid State Physics, University of Tokyo, and NSF award DMR-2004074.
	K.K. acknowledges the support by the Alexander von Humboldt Foundation.
\end{acknowledgments}

\appendix
\section{Analysis of the spectrum at the lowest measured temperature}\label{appendix:lowTspectrum}
Figure~\ref{fig:lowT_spectrum} in Section~\ref{subSec_SPC} shows the $^{13}$C NMR spectra of $\kappa$-Hg-Cl at 289.9 K and 3.92 K [identical to the data plotted in Fig.~\ref{fig:spectra}(a)], and of $\theta$-RbZn at 2.00 K~\cite{PRB-2000-Miyagawa}.
As discussed in the main text, the spectrum of $\kappa$-Hg-Cl at 289.9 K consists of two spectral lines split by the nuclear dipole interaction between neighboring $^{13}$C nuclei in ET, separated by 7.85 kHz.
At the low temperature of 3.92 K in the spin-singlet state, each spectral line of the Pake doublet observed at 289.9 K broadens into a spectrum consisting of roughly two peaks.
The total spectrum is well fitted by a four-Lorentzian function composed of two pairs of Pake doublets with center positions differing by 34 ppm.
At low temperatures in the spin-singlet state, the spin shift becomes negligible due to the decrease in magnetic susceptibility, so that the spectral shift is governed by the chemical shift. Under these conditions, the spectral distribution reflects the molecular valence-dependent charge density.
The relation between the molecular valence at ET site $i$, $\rho_{i}$, and the principal components of chemical-shift tensor, $\sigma_{i}$ was reported for $\theta$-(ET)$_{2}$I$_{3}$ as follows~\cite{Hirata_Dthesis}:
\begin{align}
		\sigma_{i}^{xx} &= 130.8\rho_{i} + 46.2\;\; \textrm{(ppm)},\notag\\
		\sigma_{i}^{yy} &= 30.4\rho_{i} + 159.0\;\; \textrm{(ppm)},\notag\\
		\sigma_{i}^{zz} &= -12.1\rho_{i} + 61.9\;\; \textrm{(ppm)}.\notag
\end{align}
Using this relation, we obtain the following relation between $\sigma_{i}$ and $\rho_{i}$ for $\kappa$-Hg-Cl:
\begin{equation}
	\sigma_{i} = 120.5 \rho_{i} + 53.3  \;\; \textrm{(ppm)}.\notag
\end{equation}
Applying the above equation, the 34 ppm difference in the shifts of the two spectra corresponds to a charge difference of $0.28e^{+}$. This value is notably smaller than the $0.7e^{+}$ in $\theta$-RbZn~\cite{PRB-2000-Miyagawa}, as indicated by the differences between $\kappa$-Hg-Cl and $\theta$-RbZn in the spectral shifts of the two Pake doublets in the spin-singlet state (Fig.~\ref{fig:lowT_spectrum}).
For reference, using a six-Lorenzian function composed of three pairs of Pake doublets to fit to the 3.92 K spectrum, we obtained the peak positions in the left part of the spectrum are 19.8, 31.6, and 61.4 ppm, which result in a charge disproportionation of 0.70e$^{+}$ : 0.45e$^{+}$ : 0.35e$^{+}$ with corresponding intensities of 1.7 : 1 : 2. 

\section{Recovery curves of $^{13}$C nuclear magnetization}\label{appendix:relaxation}
\begin{figure*}
	\includegraphics{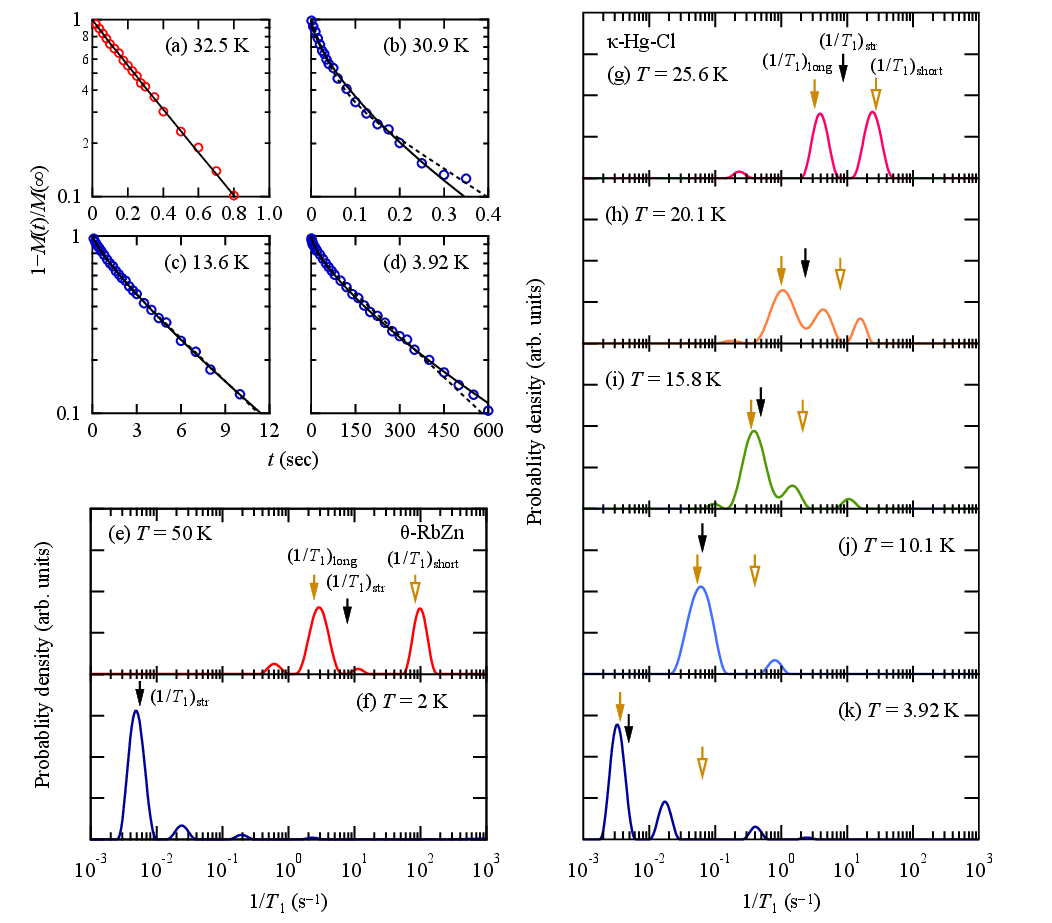}
	\caption{\label{fig:relaxation_ILT}Recovery curves of $^{13}$C nuclear magnetization at (a) 32.5,  (b) 30.9, (c) 13.6 and (d) 3.92 K. The solid and broken curves are the fits of the stretched and double exponential functions, respectively. (e--k) The probability densities of $1/T_{1}$, $P(T_{1})$ obtained from the ILT analysis using the recovery curves of $^{13}$C nuclear magnetization for $\theta$-RbZn~\cite{PRB-2000-Miyagawa} at (e) 50 K (paramagnetic state) and (f) 2 K (spin-singlet state), for $\kappa$-Hg-Cl at (g) 25.6, (h) 20.1, (i) 15.8, (j) 10.1 and (k) 3.92 K. The values of $1/T_{1}$ obtained by the stretched-exponential and double-exponential fits are indicated by arrows.}
\end{figure*}

Figures~\ref{fig:relaxation_ILT}(a)--(d) show the recovery curves of $^{13}$C nuclear magnetization acquired at several temperatures.

In the metal phase above $T_{\textrm{CO}} $ of 31 K, for example, at 32.5 K [Fig.~\ref{fig:relaxation_ILT}(a)], the best fit of the stretched exponential function to the recovery curve gave the $\beta$ value close to 1 [see also Fig.~\ref{fig:T1_CO}(b)], indicating a homogeneous relaxation.
On cooling just below $T_{\textrm{CO}}$, 30.9 K [Fig.~\ref{fig:relaxation_ILT}(b)], the recovery curve showed a bending, which gives a decrease in the $\beta$ value to 0.6--0.7. 
When we fitted it by the double exponential function, the fraction of the short- and long-$T_{1}$ components were nearly 0.5, being consistent with the occurrence of charge ordering, in which the charge-rich and charge-poor sites show short and long $T_{1}$'s, respectively. 
On further cooling, the bending in the recovery curve became more blurred [Fig.~\ref{fig:relaxation_ILT}(c)], which gives a moderate increase in $\beta$ [Fig.~\ref{fig:T1_CO}(b)].
The increase of $\beta$ as decreasing the temperature down to approximately 10 K below 25 K [Fig.~\ref{fig:T1_CO}(b)] is presumably due to the effect of spin-spin relaxation between $^{13}$C nuclei working to average the $T_{1}$'s of the charge-rich and charge-poor sites, which becomes more effective as $T_{1}$ becomes longer.
At the same time, the value of $A$ obtained from the double exponential fits decreases on cooling below 25 K.
Below 10 K, the $\beta$ value turns to a decrease [the bending of the relaxation curve gets somewhat stronger as seen in Fig.~\ref{fig:relaxation_ILT}(d)], suggesting that another relaxation component showed up, which is attributed to the orphan spins out of the spin singlet sea.

Recently, a method based on the inverse Laplace transform (ILT) has been employed to extract the NMR relaxation rate $1/T_{1}$ of electronic systems exhibiting inhomogeneous relaxation~\cite{PRB-2020-Singer,NatPhys-2021-Wang}.
Here, we applied the ILT method to analyze the $T_{1}$ relaxation of $^{13}$C nuclear magnetization in $\kappa$-Hg-Cl, as well as in the charge-ordered insulator $\theta$-RbZn~\cite{PRB-2000-Miyagawa}.
The experimental data for the saturation recovery curve for $^{13}$C nuclear magnetization, $M(t)$, is expressed as
\begin{equation}
	M(t) = \sum_{j=1}^{m}\left(1 - e^{-t/T_{1j}} \right)P(1/T_{1j})\notag
\end{equation}
where $P(1/T_{1})$ denotes the probability density function of $1/T_{1}$, which satisfying $\sum_{j=1}^{m}P(1/T_{1j}) = M_{0}$ with the saturated magnetization $M_{0}$, and $m = 300$ was chosen as the number of logarithmically spaced points between $10^{-3}$ and $10^{3}$ in the $P(1/T_{1})$ distribution.
We performed the ILT of the experimental recovery curve, $M(t_{i})$, at a set of times $t_{i}$, using the Tikhonov regularization combined with non-negative least squares to obtain $P(1/T_{1})$.
The objective function for Tikhonov regularization is:
\begin{equation}
	\arg\min_{\mathbf{P} \ge 0} \left\{ 
	\big\| \mathbf{K} \mathbf{P} - \mathbf{M} \big\|_2^2 
	+ \lambda \big\| \mathbf{L} \mathbf{P} \big\|_2^2 
	\right\},\notag
\end{equation}
where $\mathbf{P}$ denotes the discretized distribution of $P(1/T_1)$, $\mathbf{K}$ is the kernel matrix defined by $K_{ij} = \exp(-t_i /T_{1j})$, and $\mathbf{M}$ is the vector form of $M(t_{i})$. 
The matrix $\mathbf{L}$ represents the discrete second-derivative operator~\cite{JJMR-2011-Day}. 
Figures~\ref{fig:relaxation_ILT}(e) and (f) show the results for $\theta$-RbZn with $\lambda = 1$. 
In the paramagnetic charge-ordered state at 50 K, two peaks appear near the $1/T_1$ values obtained from the double-exponential fit of $M(t)$ reported in previous paper~\cite{PRB-2000-Miyagawa} [Fig.~\ref{fig:relaxation_ILT}(e)]. 
In the spin-singlet state at 2 K, the two components observed at 50 K are averaged due to the $T_{2}$ process~\cite{PRB-2000-Miyagawa}. 
The position of the main peak in $P(1/T_{1})$ in the ILT result obtained using $M(t_{i})$ at 2 K agrees with the previously reported $1/T_{1}$ values.  
Figures~\ref{fig:relaxation_ILT}(g)--(k) show the results for $\kappa$-Hg-Cl with $\lambda = 1$. 
Although the shape of the distribution function varies depending on the value of $\lambda$ when $\lambda$ is varied in the range between $10^{-4}$ and $10^{3}$, the first moment of $P(1/T_{1})$ remains nearly unchanged and  is a similar value to the $1/T_1$ value obtained from the stretched exponential fit, qualitatively reproducing its temperature dependence.

\section{Evaluation of Enhancement factor in the Korringa relation}\label{appendix:Korringa}
We calculated the enhancement factor $K(\alpha)$ in the Korringa relation, $(1/T_{1}T)K^{-2}(\hbar/4\pi k_{\textrm{B}})(\gamma_{e}/\gamma_{n}) = \zeta(\theta,\varphi)K(\alpha)$ where the geometrical factor $\zeta(\theta,\varphi)$ is 
\begin{widetext}
	\begin{equation}
			\zeta(\theta,\varphi) = \frac{a_{xx}^{2}(\sin^{2}\varphi + \cos^{2}\theta \cos^{2}\varphi)+a_{yy}^{2}(\cos^{2}\varphi + \cos^{2}\theta \sin^{2}\varphi) + a_{zz}^{2}\sin^{2}\theta}{2[a_{xx}\sin^{2}\theta\cos^{2}\varphi + a_{yy}\sin^{2}\theta \sin^{2}\varphi + a_{zz}\cos^{2}\theta]^{2}}
		\end{equation}
\end{widetext}
with the principal values of the hyperfine coupling tensor $(a_{xx}, a_{yy}, a_{zz})$.
The angles $\theta$ and $\varphi$ are the angle between the external field, $\bm{H_{0}}$, and the molecular principle $z$ axis and the polar angle of $\bm{H_{0}}$ measured from $x$ axis, respectively [Fig.~\ref{fig:structure}(d)].
As the hyperfine coupling tensor for $\kappa$-Hg-Cl was not known, we used reported values of $(a_{xx}, a_{yy}, a_{zz})$ for other ET-based compounds.
The calculated values of $\zeta(\theta, \varphi)$ and $K(\alpha)$ are summarized in Table~\ref{tab:hyperfine}.
\begin{table*}
	\caption{\label{tab:hyperfine} The principal values of the hyperfine coupling tensor reported in the literature and the values of $\zeta(\theta,\varphi)$ and $K(\alpha)$ calculated, using each principal values, for $\kappa$-Hg-Cl in an magnetic field applied parallel to $a$-axis. The $(a_{xx},a_{yy},a_{zz})$ for inner and outer $^{13}$C sites in $\kappa$-(ET)$_{2}$Cu[(CN)$_{2}$]Br were originally reported to be $(-57, -135, 423)$ ppm and $(55, -33, 728)$ ppm, respectively, in Ref.~\cite{PRB-1995-Soto}. We converted them to kOe/$\mu_{\textrm{B}}$ units using $4\times 10^{-4}$ emu/mol f.u. for the magnetic susceptibility in the metallic state~\cite{ChemRev-2004-Miyagawa}.}
	\begin{ruledtabular}
		\begin{tabular}{ccccccc}
			& & $a_{xx}$ (kOe/$\mu_{\textrm{B}}$) & $a_{yy}$ (kOe/$\mu_{\textrm{B}}$) & $a_{zz}$ (kOe/$\mu_{\textrm{B}}$) & $\zeta(\theta,\varphi)$ & $K(\alpha)$
			\\
			\hline
			$\theta$-(ET)$_{2}$I$_{3}$~\cite{PRB-2012-Hirata} & & -1.1 & -0.9 & 5.6 & 19.1 & 34.9\\
			\hline
			$\alpha$-(ET)$_{2}$I$_{3}$~\cite{PRB-2011-Hirata} & site A & -1.11 & -1.61 & 13.35 & 71.3& 9.35\\ %\cline{2-5}
			& site B & -1.26 & -2.25 & 11.91 & 61.1 & 10.9\\ %\hline
			& site C & -0.74 & -0.70 & 14.32 & 55.2& 12.1\\ %\cline{3-5}
			\hline
			$\kappa$-(ET)$_{2}$Cu[N(CN)$_{2}$]Br~\cite{PRB-1995-Soto}* & Inner site & -1.59 & -3.77 & 11.81 & 36.4 & 18.3 \\ %\cline{2-5}
			& Outer site & 1.54 & -0.92 & 20.3 & 61.4 & 10.9
		\end{tabular}
	\end{ruledtabular}
\end{table*}
The $K(\alpha)$ values were nearly 10 except for the case with the hyperfine coupling tensor for $\theta$-(ET)$_{2}$I$_{3}$ in use, which has smaller $a_{zz}$ than the others.
In $\kappa$-(ET)$_{2}$\textit{X} family, 
$K(\alpha)$ is one order of magnitude larger than unity, which indicates the presence of antiferromagnetic fluctuations.

\section{Intensity plot and $1/T_{2}$}\label{appendix:intensityT2}
\begin{figure}
	\includegraphics{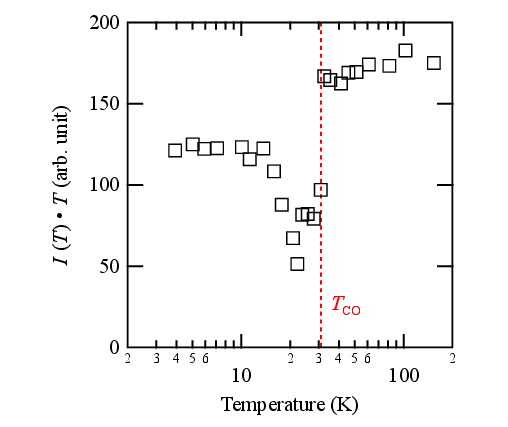}
	\caption{\label{fig:Intensity_plot} The temperature dependence of the integrated intensity of the solid-echo spectra multiplied by the temperature.}
\end{figure}

\begin{figure}
	\includegraphics{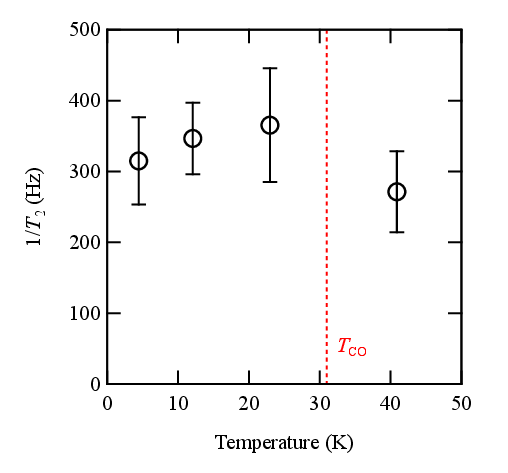}
	\caption{\label{fig:invT2} The temperature dependence of the $^{13}$C spin-spin relaxation rate $1/T_{2}$ determined by the spin echo decay.}
\end{figure}

Figure~\ref{fig:Intensity_plot} shows the temperature dependence of the integrated intensity of the solid-echo spectra of $\kappa$-Hg-Cl identical to Fig.~\ref{fig:spectra}(a). In the charge-ordered spin-gap phase, the intensity is reduced by about 20–30\% compared with that in the high-temperature metallic phase.
The solid echo focuses on the diffusion arising from the nuclear dipole interactions. When the line width broadens just below $T_{\textrm{CO}}$, the focusing is lost, leading to a decrease in signal intensity.
At lower temperatures, as the spin susceptibility decreases due to the formation of spin singlets, the signal intensity recovers.
We also measured the $^{13}$C nuclear spin-spin relaxation rate, $1/T_{2}$, by the spin echo decay (Fig.~\ref{fig:invT2}).
The temperature dependence of  $1/T_{2}$ shows no significant increase across $T_{\textrm{CO}}$.
Therefore, the loss of signal intensity across $T_{\textrm{CO}}$ is attributed to defocusing of the solid echo, rather than the wipe-out effect.

%\nocite{*}

%\bibliography{bibliography}% Produces the bibliography via BibTeX.

%

\end{document}